\documentclass[pra,aps,amsmath,amssymb,showkeys,showpacs]{revtex4}

\newcommand{\pq}{{\mathsf{p}}}
\newcommand{\qp}{{\mathsf{q}}}
\newcommand{\rp}{{\mathsf{r}}}
\newcommand{\xp}{{\mathsf{x}}}
\newcommand{\yp}{{\mathsf{y}}}
\newcommand{\cp}{{\mathsf{c}}}
\newcommand{\wqp}{{\mathsf{\tilde{q}}}}
\newcommand{\wrp}{{\mathsf{\tilde{r}}}}
\newcommand{\hh}{{\mathcal{H}}}
\newcommand{\pen}{\openone}
\newcommand{\eas}{E_{\alpha}^{(s)}}
\newcommand{\ebt}{E_{\beta}^{(t)}}
\newcommand{\nn}{\boldsymbol{n}}
\newcommand{\ma}{\boldsymbol{a}}
\newcommand{\tn}{\boldsymbol{\theta}}
\newcommand{\roh}{\boldsymbol{\rho}}
\newcommand{\tr}{{\rm{tr}}}
\newcommand{\ttm}{{\mathsf{T}}}
\newcommand{\mc}{{\mathcal{M}}}
\newcommand{\nc}{{\mathcal{N}}}
\newcommand{\ac}{{\mathcal{A}}}
\newcommand{\bc}{{\mathcal{B}}}
\newcommand{\mm}{{\mathsf{M}}}
\newcommand{\nm}{{\mathsf{N}}}
\newcommand{\am}{{\mathsf{A}}}
\newcommand{\um}{{\mathsf{U}}}

\newcommand{\vm}{{\mathsf{V}}}
\newcommand{\ax}{{\mathsf{X}}}
\newcommand{\bn}{{\mathsf{B}}}
\newcommand{\pim}{{\mathsf{\Pi}}}
\newcommand{\vata}{\vartheta}
\newcommand{\wip}{\widetilde{P}}
\newcommand{\wiw}{\widetilde{W}}
\newcommand{\tip}{\widetilde{p}}
\newcommand{\tiw}{\widetilde{w}}
\newcommand{\tir}{\widetilde{r}}
\newcommand{\tiq}{\widetilde{q}}

\begin{document}
\clearpage
\preprint{}

\title{Number-phase uncertainty relations in terms of generalized entropies}
\author{Alexey E. Rastegin}
\affiliation{Department of Theoretical Physics, Irkutsk State University,
Gagarin Bv. 20, Irkutsk 664003, Russia}

\begin{abstract}
Number-phase uncertainty
relations are formulated in terms of unified entropies which form
a family of two-parametric extensions of the Shannon entropy. For
two generalized measurements, unified-entropy uncertainty
relations are given in both the state-dependent and
state-independent forms. A few examples are discussed as well.
Using the Pegg--Barnett formalism and the Riesz theorem, we obtain
a nontrivial inequality between norm-like functionals of generated
probability distributions in finite dimensions. The principal
point is that we take the infinite-dimensional limit right for
this inequality. Hence number-phase uncertainty relations with
finite phase resolutions are expressed in terms of the unified
entropies. Especially important case of multiphoton coherent
states is separately considered. We also give some entropic bounds
in which the corresponding integrals of probability density
functions are involved.
\end{abstract}

\maketitle

\pagenumbering{arabic}
\setcounter{page}{1}

\section{Introduction}

Since celebrate Heisenberg's result \cite{heisenberg} had been
published, many forms of uncertainty relations were proposed
\cite{hall99,lahti}. A new interest was inspired by recent
advances in use of quantum information. Entropy plays a central
role in statistical physics and information theory. Entropic
measures have found use in various topics including a
quantification of uncertainty in quantum measurements
\cite{pp96,ww10}. Within the most known approach by Robertson
\cite{robert}, some doubts have been observed
\cite{deutsch,maass}. An alternate way is to express the
uncertainty principle by means of information-theoretic terms.
Although the Shannon entropy is of great importance, many
generalizations are found to be useful. Both the R\'{e}nyi
\cite{renyi61} and Tsallis entropies \cite{tsallis} have widely
been adopted for interdisciplinary applications. The authors of
the paper \cite{hey06} proposed the notion of unified entropy
which includes the above entropies as particular cases. Properties
of unified entropies were examined in both the classical and
quantum regimes \cite{hey06,rastjst}.

The first relation in terms of the Shannon entropies for the
position-momentum pair was derived by Hirschman \cite{hirs}. An
improvement of his result has been stated in Ref. \cite{beck} (see
also Ref. \cite{birula1}). Mamojka \cite{mamojka} and Deutsch
\cite{deutsch} initiated a discussion of the problem in general
form and obtained particular results. An improvement of Deutsch's
entropic bound \cite{deutsch} had been conjectured in Ref.
\cite{kraus87} and proved with use of Riesz's theorem in Ref.
\cite{maass}. In Ref. \cite{rastijtp}, we have discussed
two-measurement entropic bounds that cannot be derived on the base
of Riesz's theorem. The time-energy case \cite{hall08} and
tomographic processes \cite{mmanko10} were considered within
entropic approach. Cryptography applications of entropic
uncertainty relations are discussed in Refs. \cite{dmg07,BCCRR09}.
Entropic uncertainty relations have also been given for more than
two measurements. Entropic bounds for $(N+1)$ complementary
observables in $N$-dimensional Hilbert space were derived in Refs.
\cite{ivan92,jsan93}. New results in this issue have been reported
in Ref. \cite{aamb10}. For arbitrary number of binary observables,
a nearly optimal relation for the collision entropy was given in
Ref. \cite{ww08}.

The problem of constructing a Hermitian operator of phase has a long
history (see the review \cite{lynch} and references therein).
Different ways to measure a quantum phase uncertainty are compared
in Refs. \cite{bfs93,mjwh93}. Both the well-defined Hermitian
operator of phase and corresponding number-phase relation of
Robertson type have been obtained within the Pegg--Barnett
formalism \cite{BP89,PB89} (see also section 4.3 in Ref.
\cite{radmore}). Entropic relations for the number-phase pair in
terms of the Shannon entropies were obtained in Refs.
\cite{abe92,gvb95,jos01}. Using some ideas of Ref. \cite{gvb95},
entropic relations of ''number-phase'' type have been posed for
solvable quantum systems with discrete spectra \cite{hatami}.
Uncertainty relations for the number and annihilation operators
have been considered in Refs. \cite{lanz,rast105}.

In the present paper, we formulate number-phase uncertainty
relations in terms of unified entropies. The phase operator is
approached within the Pegg--Barnett formalism \cite{BP89,PB89}.
The paper is organized as follows. In Section \ref{sc1}, the
required preliminary material is presented. For two generalized
measurements, the entropic uncertainty relations in terms of
unified entropies are derived in Section \ref{sc2}. Entropic
bounds of both state-dependent and state-independent forms are
simultaneously treated. Several examples of interest are
considered in Section \ref{sc3}. One includes the cases of
complementary observables for $N$-level system, angle and angular
momentum, and extremal unravelings of quantum channels. In Section
\ref{sc4}, the developed method is appropriately modified. Hence
number-phase uncertainty relations in terms of unified entropies
are immediately derived. Incidentally, the case of canonically
conjugate variables is mentioned. Section \ref{sc5} concludes the
paper with a summary of results.

\section{Definitions and background}\label{sc1}

In this section, we introduce some terms and conventions that will
be used through the text. First, we recall operators and states
commonly used in quantum optics. An outline of the Pegg--Barnett
formalism is given as well. Finally, the utilized entropic
measures are considered.

\subsection{Operators and states. Elements of the Pegg--Barnett formalism}\label{s11}

It is well known that the energy eigenstates for a
single field mode are the number states  analogous to those for
the harmonic oscillator \cite{radmore}. For given mode, the
annihilation and creation operators $\ma$ and $\ma^{\dagger}$
satisfy the commutation rule $[\ma,\ma^{\dagger}]=\pen$, where
$\pen$ is the identity operator. By $|n\rangle$ we denote the
normalized eigenstate of the number operator
$\nn=\ma^{\dagger}\ma$, so that $\nn|n\rangle=n|n\rangle$ with
integer $n\geq0$. These states form an orthonormal basis in the
Hilbert space. Under the action of the annihilation and creation
operators, the numbers states are transformed as
$\ma|n\rangle=n^{1/2}|n-1\rangle$ and
$\ma^{\dagger}|n\rangle=(n+1)^{1/2}|n+1\rangle$. In thermal
equilibrium, the state of a system is described by the density
operator corresponding to the grand canonical ensemble. For a
single field mode with frequency $\omega$ in a thermal state, the
density operator is diagonal in the basis $\{|n\rangle\}$ with the
probabilities that follow the Bose--Einstein statistics
\cite{radmore}. Coherent states form another class which is
especially important in quantum optics. For a complex number
$z=|z|e^{i\phi}$, we write coherent state as \cite{walls}
\begin{equation}
|z\rangle=\exp\bigl(-|z|^{2}/2\bigr)\sum_{n=0}^{\infty} \frac{z^{n}}{\sqrt{n!}}{\>}|n\rangle
\ . \label{gams}
\end{equation}
The state $|z\rangle$ obeys $\ma|z\rangle=z|z\rangle$ and
$\langle{z}|\ma^{\dagger}=\langle{z}|z^{*}$, i.e. coherent states
are {\it right} eigenstates of the annihilation operator and {\it
left} eigenstates of the creation one. Coherent states are not
mutually orthogonal, but do satisfy the completeness relation
$(1/\pi)\int{d^{2}}z{\,}|z\rangle\langle{z}|=\pen$, where the
double integral is taken over the whole complex plane
\cite{walls,KS85}. Coherent states have many nice properties and
various applications (see, e.g., a collection of papers in Ref.
\cite{KS85} and Ref. \cite{sanders} with a focus on entangled
coherent states). We only recall that the number of photons in a
single-mode coherent state $|z\rangle$ has the Poisson
distribution \cite{lkn87}. Hence, the mean number of photons and
the photon number variance are both equal to $|z|^{2}$.

In order to represent the phase operator, we first consider
operators and states in finite-dimensional state spaces. Let
$\{|n\rangle\}$ be an orthonormal basis in $(N+1)$-dimensional
space $\hh_{N+1}$. The number operator is decomposed as
\begin{equation}
\nn_{N+1}=\sum\nolimits_{n=0}^{N}n|n\rangle\langle{n}|
\ . \label{nuop}
\end{equation}
We introduce normalized states
\begin{equation}
|\theta_{m}\rangle=\frac{1}{\sqrt{N+1}}\sum\nolimits_{n=0}^{N}e^{in\theta_{m}}|n\rangle
\ , \label{tmb}
\end{equation}
where $\theta_{m}=\theta_0+2\pi{m}/(N+1)$ and $m=0,1,2,\ldots,N$.
It is clear that the vectors $|\theta_{m}\rangle$ form another
orthonormal basis in $\hh_{N+1}$. Note that the bases
$\{|n\rangle\}$ and $\{|\theta_m\rangle\}$ are mutually unbiased.
The phase operator is defined as
\begin{equation}
\tn_{N+1}=\sum\nolimits_{m=0}^{N}\theta_{m}|\theta_{m}\rangle\langle\theta_{m}|
\ . \label{phop}
\end{equation}
Within the considered formalism, quantities of interest are first
calculated for finite $N$ and then taken in the limit
$N\to\infty$. In particular, the moments of the phase operator can
be obtained in this way (for details, see section 4.3 in
\cite{radmore}). For physical states, however, there is a more
straightforward approach. Following the Pegg--Barnett formalism
\cite{BP89,PB89}, we introduce the phase probability density
function by
\begin{equation}
P(\theta)=\underset{N\to\infty}{\lim}
\frac{N+1}{2\pi}{\>}\langle\theta|\roh|\theta\rangle=\frac{1}{2\pi}\sum_{m,n=0}^{\infty}
\langle{m}|\roh|n\rangle\exp\bigl[i(n-m)\theta\bigr]
\ . \label{ppdf}
\end{equation}
Converting the summation into a Riemann--Darboux integral, we then
write the moment of order $\nu$ as
\begin{equation}
\tr\bigl(\tn_{N+1}^{{\>}\nu}\roh\bigr)\underset{N\to\infty}{\longrightarrow}
\int_{\theta_{0}}^{\theta_{0}+2\pi}{\theta^{\nu}P(\theta){\,}d\theta}
\ . \label{numom}
\end{equation}
If the density operator $\roh$ is diagonal with respect to the
basis $\{|n\rangle\}$ then the right-hand side of Eq. (\ref{ppdf})
gives $P(\theta)=(2\pi)^{-1}$. In particular, this formula holds
for thermal states of a single mode. Since the function
$P(\theta)$ is constant here, these are states of completely
unknown phase. For coherent states, the series in Eq. (\ref{ppdf})
is difficult to evaluate. However, for large values of $|z|$ a
useful expression exists. Namely, for $|z|\gg1$ the corresponding
density is commonly approximated by the Gaussian distribution
\cite{radmore}
\begin{equation}
P(\theta)\simeq\left(\frac{2{\,}|z|^{2}}{\pi}\right)^{1/2}
\exp\Bigl\{-2{\,}|z|^{2}(\theta-\phi)^{2}\Bigr\}=:\wip(\theta)
\ . \label{gampa}
\end{equation}
Since the Gaussian distribution (\ref{gampa}) has a narrow peak at
$\theta=\phi$, such states $|z\rangle$ are physical states of
almost determined phase. Note that no physical states with
completely determined phase exist. Since the photon number
variance is very large here, the Poisson number distribution is
very spreading and can also be approximated by the Gaussian
distribution \cite{radmore,lkn87}
\begin{equation}
\wiw(n):=\left(\frac{1}{2\pi|z|^{2}}\right)^{1/2}
\exp\left\{-\frac{\left(n-|z|^{2}\right)^{2}}{2{\,}|z|^{2}}\right\}
\ , \label{nampa}
\end{equation}
where the $n$ is now treated as continuous. This distribution is
normalized in the sense that
$\int_{-\infty}^{+\infty}\wiw(n){\,}dn=1$. Here we must stress the
following. The expressions (\ref{gampa}) and (\ref{nampa}) are
assumed to be used as convenient approximations only for
$|z|\gg1$. In this case, the maximum of $\wiw(n)$ at $n=|z|^{2}$
is equal to $\left(\sqrt{2\pi}{\,}|z|\right)^{-1}\ll1$. So values
of this function are negligible for negative $n$ as well. It is
important for our aims that the distributions (\ref{gampa}) and
(\ref{nampa}) can be related via the Fourier transform. Namely, we have
\begin{equation}
\tip(\xi)=\frac{1}{\sqrt{2\pi}}\int\nolimits_{-\infty}^{+\infty}e^{i\xi\kappa}{\,}\tiw(\kappa){\,}d\kappa
\ , \qquad
\tiw(\kappa)=\frac{1}{\sqrt{2\pi}}\int\nolimits_{-\infty}^{+\infty}e^{-i\kappa\xi}{\,}\tip(\xi){\,}d\xi
\ , \label{tipw}
\end{equation}
where $\wip(\theta)=\tip(\xi)^{2}$ and
$\wiw(n)=\tiw(\kappa)^{2}$ in terms of the variables
$\xi=\theta-\phi$ and $\kappa=n-|z|^{2}$.

\subsection{Generalized entropies and their forms}\label{s13}

The concept of entropy is a key tool in information theory. In
addition to the Shannon entropy, which is fundamental, other
entropic measures were found to be useful. Among them, the
R\'{e}nyi and Tsallis entropic functionals are very important
\cite{bengtsson}. For given probability distribution
$\pq=\{p_{n}\}$, its R\'{e}nyi $\alpha$-entropy is defined as
\cite{renyi61}
\begin{equation}
R_{\alpha}(\pq):=\frac{1}{1-\alpha}{\ }\ln\left(\sum\nolimits_{n} p_{n}^{\alpha}\right)
\ , \label{rpdf}
\end{equation}
where $\alpha>0$ and $\alpha\neq1$. This quantity is a
non-increasing function of $\alpha$ \cite{renyi61}. For other
properties, related to the parametric dependence, see Ref.
\cite{zycz}. The Renyi entropy of order $\alpha=2$ is also known
as the collision entropy \cite{ww10,ww08}. The notion of Tsallis
entropy is widely used in non-extensive statistical mechanics
\cite{gmt}. The non-extensive entropy of positive degree
$\alpha\neq1$ is defined as \cite{tsallis}
\begin{equation}
H_{\alpha}(\pq):=\frac{1}{1-\alpha}{\,}
\left(\sum\nolimits_{n} p_{n}^{\alpha}-1\right)=
-\sum\nolimits_{n} p_{n}^{\alpha}\ln_{\alpha}(p_{n})
\ , \label{tsent}
\end{equation}
where $\ln_{\alpha}(\xi):=\bigl(\xi^{1-\alpha}-1\bigr)/(1-\alpha)$
is the $\alpha$-logarithm. Its inverse function is the
$\alpha$-exponential function
\begin{equation}
\exp_{\alpha}(\xi):=\Bigl(1+(1-\alpha){\,}\xi\Bigr)_{+}^{1/(1-\alpha)}
\ , \label{qexf}
\end{equation}
where $(\xi)_{+}:=\max\{0,\xi\}$. For $\alpha\to1$, the
$\alpha$-logarithm and the $\alpha$-exponential are
respectively reduced to the usual ones. With slightly other
factor, the entropic functional (\ref{tsent}) was derived from
several axioms by Havrda and Charv\'{a}t \cite{HC67}. The Tsallis
entropy of degree $\alpha=2$ is related to the so-called degree of
certainty involved in complementarity relations for mutually
unbiased observables \cite{diaz}. The R\'{e}nyi and Tsallis
entropies are fruitful one-parametric generalizations of the
Shannon entropy. In the limit $\alpha\to1$, the entropies
(\ref{rpdf}) and (\ref{tsent}) both recover the standard Shannon
entropy $S(\pq)=-\sum_{n}p_{n}\ln{p}_{n}$. In Ref. \cite{hey06},
Hu and Ye proposed the class of more general extensions of the
Shannon entropy. For positive $\alpha\neq1$ and $s\neq0$, the
unified $(\alpha,s)$-entropy of probability distribution
$\pq=\{p_{n}\}$ is defined by
\begin{equation}
\eas(\pq):=\frac{1}{(1-\alpha){\,}s}{\,}\left[\left(\sum\nolimits_{n} p_{n}^{\alpha}\right)^{{\!}s}-1\right]
\ . \label{undef}
\end{equation}
This two-parametric entropic functional includes the entropy
(\ref{tsent}) as the partial case $s=1$ and the R\'{e}nyi entropy
(\ref{rpdf}) as the limiting case $s\to0$ \cite{hey06}. In quantum
regime, this entropic form leads to a two-parametric extension of
the von Neumann entropy. The quantum unified entropies enjoy many
properties similarly to the von Neumann entropy
\cite{hey06,rastjst}.

The above expressions for entropies can be rewritten in terms of
norm-like functionals. For $b\geq1$, the $l_b$ norm of vector
$\xp=\{x_{n}\}$ is written as \cite{watrous1}
\begin{equation}
\|\xp\|_b:=\left(\sum\nolimits_{n}|x_{n}|^b\right)^{1/b}
\ . \label{xbnm}
\end{equation}
These norms are convenient for studying uncertainty relations in
the context of information processing \cite{vsw09}. For
$b=\infty$, this norm is defined as
$\|\xp\|_{\infty}:=\max_{n}|x_{n}|$. For probability distribution
$\pq=\{p_{n}\}$ and $\beta>0$, we then introduce a norm-like
$\beta$-functional
\begin{equation}
\|\pq\|_{\beta}:=\left(\sum\nolimits_{n}p_{n}^{\beta}\right)^{1/\beta}
\ . \label{pbnf}
\end{equation}
This expression is actually a norm for $\beta\geq1$, but we will
also use it for $\beta\in[1/2;1)$. In terms of these functionals,
the entropy (\ref{undef}) is rewritten as
\begin{equation}
\eas(\pq)=\frac{\|\pq\|_{\alpha}^{\alpha{s}}-1}{(1-\alpha){\,}s}
\ , \label{unfd}
\end{equation}
including
$H_{\alpha}(\pq)=\bigl(\|\pq\|_{\alpha}^{\alpha}-1\bigr)/(1-\alpha)$
for $s=1$ and
$R_{\alpha}(\pq)=\alpha(1-\alpha)^{-1}\ln\|\pq\|_{\alpha}$ for
$s=0$. When a probability distribution is continuous, we merely
replace each sum with proper integral of the probability density
function.

Utility of entropic approach for measuring uncertainty in quantum
measurement is widely discussed (see the works
\cite{ww10,deutsch,maass,rast102,brud11} and references therein).
The only point we should note about entropic measures for a
continuous distribution is that they can be used in two different
forms. The first form with clear operational meaning is written
for some partition of the interval, in which continuous variable
does range. For the phase, we consider a partition
$\{\vata_{m}\}$ of the interval $[0;2\pi]$. The probability to find
value of the phase variable in the $m$th bin is written as
\begin{equation}
r_{m}=\int\nolimits_{\vata_{m}}^{\vata_{m+1}} P(\theta){\,}d\theta
\ . \label{rbin}
\end{equation}
For probability vector $\rp=\{r_{m}\}$, the corresponding
$(\alpha,s)$-entropy $\eas(\rp)$ is then defined in line with Eq.
(\ref{undef}). As a rule, the bins are all of the same size which
characterizes a resolution of measurement. Further, we can
directly introduce entropic measures
\begin{align}
\eas(P)&=\frac{1}{(1-\alpha){\,}s}
{\>}\left\{\left(\int\nolimits_{0}^{2\pi}P(\theta)^{\alpha}{\,}d\theta\right)^{\!s}-1\right\}
\qquad (s\neq0)
\ , \label{cddfs} \\
R_{\alpha}(P)&=\frac{1}{1-\alpha}{\ } \ln\left(\int\nolimits_{0}^{2\pi}P(\theta)^{\alpha}{\,}d\theta\right)
\ . \label{cddf0}
\end{align}
For $\alpha=1$, the Shannon entropy is
$S(P)=-\int\nolimits_{0}^{2\pi}P(\theta)\ln{P(\theta)}{\,}d\theta$.
The latter measure is used in number-phase uncertainty relations
given in Refs. \cite{abe92,gvb95,jos01}. With the entropies
(\ref{cddfs}) and (\ref{cddf0}), however, we have observed some
doubts. The discrete $(\alpha,s)$-entropy (\ref{undef}) is always
positive, and vanish only when all the probabilities, except one,
are zero. By $\|\rp\|_{1}=1$, we actually have
$\|\rp\|_{\alpha}\leq1\leq\|\rp\|_{\beta}$ for $\alpha>1>\beta$.
It is not the case for continuous distributions. In general, the
normalization $\|P\|_{1}=1$ does not provide
\begin{equation}
\|P\|_{\alpha}\leq1\leq\|P\|_{\beta}
\label{pacon}
\end{equation}
for $\alpha>1>\beta$. If the condition (\ref{pacon}) fails then
the quantities (\ref{cddfs}) and (\ref{cddf0}) become
negative-valued. We do not treat such functionals for expressing
an uncertainty, though the Shannon entropy of the phase with
negative values up to $-\infty$ was noted in the literature (see,
e.g., section 2.3 of Ref. \cite{mjwh93}). Instead, we will mainly
use entropies calculated for discretized distribution with
probabilities of a kind (\ref{rbin}). The entropies (\ref{cddfs})
and (\ref{cddf0}) cannot be negative, when the probability density
function obeys the condition (\ref{pacon}). Say, this condition is
always satisfied with the density such that $P(\theta)\leq1$ for
all $\theta\in[0;2\pi]$. The class of densities with sufficiently
small variations is wide enough. For instance, such densities
stand for thermal states and, generally, for those density
matrices that are diagonal in the number eigenbasis
$\{|n\rangle\}$. In Ref. \cite{rast105} we have analyzed an
example, in which the probability density function does not exceed
one with necessity. On the other hand, coherent states $|z\rangle$
with large $|z|$, for which the formula (\ref{gampa}) takes place,
are clearly beyond the above class. Physical quantities are
usually dimensional that leads to another question for entropic
functionals expressed as integrals. Since the phase is
dimensionless, we do not enter into details and refer to the
extensive review \cite{brud11}.

\section{Entropic uncertainty relations for two generalized measurements}\label{sc2}

In this section, entropic uncertainty relations for two
generalized measurements are posed in terms of unified entropies.
First, we recall an inequality between norm-like
functionals of the generated probability distributions. This
statement is based on Riesz's theorem, which is also recalled.
Second, we find the minimum of a certain function of two variables
in the proper domain. The results are then used for deriving
unified-entropy uncertainty relations.

\subsection{An inequality based on Riesz's theorem}\label{s21}

A generalized quantum measurement is described by ''positive
operator-valued measure'' (POVM). This is a set $\mc=\{\mm_{i}\}$ of
positive semidefinite operators obeying the completeness relation
$\sum_{i}\mm_{i}=\pen$ \cite{peresq}. The standard von Neumann
measurement is represented by the set of mutually orthogonal
projectors. For given POVM $\mc=\{\mm_{i}\}$ and density
operator $\roh$, the probability of $i$th outcome is expressed as
$p_{i}=\tr(\mm_{i}\roh)$ \cite{peresq}. The Hilbert--Schmidt inner
product of operators $\am$ and $\bn$ is defined as
$\langle\am{\,},\bn\rangle_{\rm{hs}}:=\tr(\am^{\dagger}{\,}\bn)$. To
get entropic relations, we will use a version of Riesz's theorem
(see theorem 297 in the book \cite{hardy}). Consider the tuples $\xp$ and $\yp$ of
complex numbers related by a linear transformation
$\ttm$ as
\begin{equation}
y_{i}=\sum\nolimits_{j}\tau_{ij}{\>}x_{j}
\ . \label{lintrans}
\end{equation}
Let $\eta$ be maximum of $|\tau_{ij}|$, i.e.
$\eta:={\max}|\tau_{ij}|$, and let conjugate indices
$a,b\in[1;\infty]$ obey $1/a+1/b=1$. If the matrix
$\ttm=[[\tau_{ij}]]$ satisfies
$\|{\mathsf{y}}\|_{2}\leq\|{\mathsf{x}}\|_{2}$ for all $\xp$ and
$1<b<2$, then
\begin{equation}
\|\yp\|_{a}\leq{\eta}^{(2-b)/b}\|\xp\|_{b}
\ . \label{suppos}
\end{equation}
So, Riezs's theorem provides an upper estimate on the $(b,a)$-norm
of transformation $\ttm$. For more general versions, see the book
\cite{berg} and references therein. Using Eq. (\ref{suppos}), we
deduce the following \cite{rast104}. Let $\mc=\{\mm_{i}\}$ and
$\nc=\{\nm_{j}\}$ be two POVMs. For given $\roh$, the
corresponding probabilities are written as
$p_{i}=\tr(\mm_{i}\roh)$ and $q_{j}=\tr(\nm_{j}\roh)$. Then we
have \cite{rast104}
\begin{equation}
\|\pq\|_{\alpha}\leq{g}(\mc,\nc|\roh)^{2(1-\beta)/\beta}{\,}
\|\qp\|_{\beta}
\ , \label{epqs1}
\end{equation}
where $1/\alpha+1/\beta=2$, $1/2<\beta<1$, and the function
$g(\mc,\nc|\rho)$ is determined by
\begin{equation}
g(\mc,\nc|\roh):=\max\left\{(p_{i}q_{j})^{-1/2}
{\,}\bigl|\tr(\mm_{i}\nm_{j}\roh)\bigr|:{\>} p_{i}\neq0,{\,} q_{j}\neq0 \right\}
. \label{gfdef}
\end{equation}
Note that $g(\mc,\nc|\roh)=g(\nc,\mc|\roh)$ in view of
$\tr(\mm_{i}\nm_{j}\roh)=\left\langle\mm_{i}\sqrt{\roh}{\,},\nm_{j}\sqrt{\roh}\right\rangle_{\rm{hs}}$.
To approach relations of a state-independent form, we use the
inequality \cite{rast104}
\begin{equation}
g(\mc,\nc|\roh)\leq\bar{f}(\mc,\nc):=
\max\Bigl\{\bigl\|\mm_{i}^{1/2}\nm_{j}^{1/2}\bigr\|_{\infty}:
{\ }\mm_{i}\in\mc,{\,}\nm_{j}\in\nc\Bigr\}
{\>}. \label{gfmn}
\end{equation}
Here the spectral norm $\|\am\|_{\infty}$ is put as the largest
eigenvalue of operator $\sqrt{\am^{\dagger}\am}\geq{\mathbf{0}}$.
The relation (\ref{gfmn}) shows that $g(\mc,\nc|\roh)\leq1$.
Combining Eqs. (\ref{epqs1}) and (\ref{gfmn}) finally gives
\begin{equation}
\|\pq\|_{\alpha}\leq{\bar{f}}(\mc,\nc)^{2(1-\beta)/\beta}{\,}
\|\qp\|_{\beta}
\ , \label{epqs2}
\end{equation}
under the same conditions on $\alpha$ and $\beta$. In Ref.
\cite{rast104}, the inequalities (\ref{epqs1}) and (\ref{epqs2})
were used to derive entropic uncertainty relations in terms of
both the R\'{e}nyi and Tsallis entropies. We shall now extend the
method of  Ref. \cite{rast104} to the case of
$(\alpha,s)$-entropies.

\subsection{Entropic uncertainty relations in terms of $(\alpha,s)$-entropies}\label{s22}

Putting two variables $\xi=\|\pq\|_{\alpha}^{\alpha}$ and
$\zeta=\|\qp\|_{\beta}^{\beta}$, the sum of two unified entropies
reads
\begin{equation}
E_{\alpha}^{(s)}(\pq)+E_{\beta}^{(t)}(\qp)=
\frac{\xi^{s}-1}{(1-\alpha){\,}s}+\frac{\zeta^{t}-1}{(1-\beta){\,}t}=:h(\xi,\zeta)
\ . \label{dfhf}
\end{equation}
Assuming $\alpha>1>\beta$, we clearly have $\xi\leq1$ and
$\zeta\geq1$. The function $h(\xi,\zeta)$ is zero at the point
$(1,1)$. We now add Eq. (\ref{epqs1}) in the form
\begin{equation}
c{\,}\xi^{\beta/\alpha}\leq\zeta,
\qquad
c=g(\mc,\nc|\roh)^{-2(1-\beta)}
\ . \label{epqs3}
\end{equation}
Non-trivial uncertainty relations take place for the case
$g(\mc,\nc|\roh)<1$, in which $c>1$ due to $\beta<1$. Then the
curve $\zeta=c{\,}\xi^{\beta/\alpha}$ cuts off the down right
corner of the rectangle
$\bigl\{(\xi,\zeta):{\>}0\leq\xi\leq1,{\>}1\leq\zeta<+\infty\bigr\}$
and, herewith, the point $(1,1)$ (see, e.g., Fig. A1 of Ref.
\cite{rast104}). So we have arrived at a task of minimizing
$h(\xi,\zeta)$ in the domain
\begin{equation}
D:=\left\{(\xi,\zeta):{\>}0\leq\xi\leq1,{\>}1\leq\zeta<+\infty,
{\>}c{\,}\xi^{\beta/\alpha}\leq\zeta\right\}
\ . \label{dodf}
\end{equation}

\newtheorem{t31}{Lemma}
\begin{t31}\label{lemun}
Let strictly positive numbers $\alpha$ and $\beta$ obey
$1/\alpha+1/\beta=2$, and let two real numbers $s$ and $t$ obey
$st>0$. The minimum of the function $h(\xi,\zeta)$ in the domain
$D$ is equal to
\begin{equation}
\underset{D}{\min}{\>}h(\xi,\zeta)=\frac{1}{\nu}{\>}\ln_{\mu}\!\left(g^{-2\nu}\right)
\ . \label{minfn}
\end{equation}
Here the parameters $\mu$ and $\nu$ are defined as
\begin{equation}
\mu:=
\left\{
\begin{array}{cc}
\max\{\alpha,\beta\}{\>}, & s,t\in(0;+\infty) \\
\min\{\alpha,\beta\}{\>}, & s,t\in(-\infty;0)
\end{array}
\right\}
\ , \qquad
\nu:=s(\mu)
\ , \label{nupos}
\end{equation}
under the notation $s(\alpha)\equiv{s}$ and $s(\beta)\equiv{t}$.
\end{t31}

{\bf Proof.} For definiteness, we assume that $\alpha>1>\beta$. By
$(\xi_{0},1)$ with $\xi_{0}=c^{-\alpha/\beta}$, we denote the
point of intersection of the lines $\zeta=1$ and
$\zeta=c{\,}\xi^{\beta/\alpha}$. Since $h(1,1)=0$, the desired
minimum is strictly positive. In the interior of $D$, we have
\begin{equation}
\frac{\partial{h}}{\partial\xi}=\frac{\xi^{s-1}}{1-\alpha}<0  \ , \qquad
\frac{\partial{h}}{\partial\zeta}=\frac{\zeta^{t-1}}{1-\beta}>0  \ ,
\label{partf}
\end{equation}
due to $\alpha>1>\beta$. Hence the minimal value is reached on the
boundary of the domain $D$. The derivatives (\ref{partf}) show
that the function $h(\xi,\zeta)$ is decreasing in $\xi$ and
increasing in $\zeta$. On the linear boundary segments, therefore,
the minimal value is either $h(\xi_{0},1)$ or $h(1,c)$. Let us
consider the boundary segment on the curve
$\zeta=c{\,}\xi^{\beta/\alpha}$. Substituting
$\zeta=(\xi/\xi_{0})^{\beta/\alpha}$ in the expression of
$h(\xi,\zeta)$ and differentiating with respect to $\xi$, we
obtain the derivative
\begin{equation}
\frac{1}{(1-\alpha){\,}\xi}{\ }\xi^{s}+
\frac{\beta}{\alpha(1-\beta){\,}\xi}\left(\frac{\xi}{\xi_{0}}\right)^{t\beta/\alpha}=
\frac{1}{(\alpha-1){\,}\xi}\left(\left(\frac{\xi}{\xi_{0}}\right)^{t\beta/\alpha}-{\,}\xi^{s}\right)
\ . \label{deriv}
\end{equation}
Here the equality $\beta/(1-\beta)=\alpha/(\alpha-1)$ was used due
to $1/\alpha+1/\beta=2$. We shall now consider the following
cases: (i) $s>0$ and $t>0$; (ii) $s<0$ and $t<0$.

In the case (i), the powers $t\beta/\alpha$ and $s$ in the
right-hand side of (\ref{deriv}) are both strictly positive and
\begin{equation}
\xi^{s}\leq1\leq(\xi/\xi_{0})^{t\beta/\alpha}
\label{derpos}
\end{equation}
by $\xi_{0}\leq\xi\leq1$. So the derivative (\ref{deriv}) is
positive and the minimum is reached at the point $(\xi_{0},1)$. By
calculations,
\begin{equation}
h(\xi_{0},1)=\frac{\xi_{0}^{s}-1}{(1-\alpha)s}=\frac{c^{-s\alpha/\beta}-1}{(1-\alpha)s}=
\frac{g^{2s(\alpha-1)}-1}{(1-\alpha)s}
=\frac{1}{s}{\>}\ln_{\alpha}\!\left(g^{-2s}\right)
\ . \label{cas1}
\end{equation}
If $\alpha>1>\beta$ then $\mu=\alpha$ and $\nu=s$ due to the
definition (\ref{nupos}). Hence the right-hand sides of
(\ref{minfn}) and (\ref{cas1}) coincide.

In the case (ii), the powers $t\beta/\alpha$ and $s$ in the
right-hand side of (\ref{deriv}) are both strictly negative and
\begin{equation}
(\xi/\xi_{0})^{t\beta/\alpha}\leq1\leq\xi^{s}
\label{derneg}
\end{equation}
by $\xi_{0}\leq\xi\leq1$. So the derivative (\ref{deriv}) is
negative and the minimum is reached at the point $(1,c)$. By
calculations,
\begin{equation}
h(1,c)=\frac{c^t-1}{(1-\beta)t}=
\frac{g^{-2t(1-\beta)}-1}{(1-\beta)t}=\frac{1}{t}{\>}\ln_{\beta}\!\left(g^{-2t}\right)
\ . \label{cas2}
\end{equation}
If $\alpha>1>\beta$ then $\mu=\beta$ and $\nu=t$ by the definition
(\ref{nupos}). Hence the right-hand sides of (\ref{minfn}) and
(\ref{cas2}) coincide. $\blacksquare$

Applying the result of Lemma \ref{lemun}, we get both the
state-dependent and state-independent forms of uncertainty
relations in terms of $(\alpha,s)$-entropies. Note that the
right-hand side of Eq. (\ref{minfn}) tends to $-2\ln{g}$ in the
limit $\nu\to0$. Taking $s=0$ and $t=0$, we then obtain
uncertainty relations in terms of the R\'{e}nyi entropies. In
effect, these relations can immediately be derived from Eqs.
(\ref{epqs1}) and (\ref{epqs2}) by simple algebra (for details,
see Refs. \cite{rast102,rast104}). Adding the case of R\'enyi's
entropies, we have the following statement.

\newtheorem{t32}{Theorem}
\begin{t32}\label{mthm}
Let $\mc=\{\mm_{i}\}$ and $\nc=\{\nm_{j}\}$ be two POVMs, and let
$\roh$ be a density matrix. For $st>0$ and $s=t=0$, there holds
\begin{equation}
E_{\alpha}^{(s)}(\pq)+E_{\beta}^{(t)}(\qp)\geq
\frac{1}{\nu}{\>}\ln_{\mu}\Bigl\{g(\mc,\nc|\roh)^{-2\nu}\Bigr\}
\ , \label{trun}
\end{equation}
where $1/\alpha+1/\beta=2$, the function $g(\mc,\nc|\roh)$ is
given by Eq. (\ref{gfdef}) and the numbers $\mu$ and $\nu$ are
given by Eq. (\ref{nupos}). The state-independent form is obtained
by replacing $g(\mc,\nc|\roh)$ with $\bar{f}(\mc,\nc)$ defined by
(\ref{gfmn}).
\end{t32}

So, we have obtained uncertainty relations for many of the unified
entropies of generated probability distributions. The entropic
bound (\ref{trun}) is valid for generalized measurements as well
as for any mixed state of interest. Few comments on
state-independent entropic bounds must be given here. The
right-hand side of Eq. (\ref{gfmn}) has been adopted with use of
Riesz's theorem. This theorem is generally applicable to any
linear transformation, but only an upper estimate on its
$(b,a)$-norm is provided in this way. As a rule, it is very
difficult to find exactly values of such norms. However, these
exact values are known for some important cases, including the
Fourier transform in both the discrete and continuous varieties.
In the former, the exact values are given by the Young-Hausdorff
inequalities (see, e.g., point (2.25) of chapter XII in
\cite{zygm2}); in the latter, the ones have been found
by Beckner \cite{beck}. When the exact estimates are known, we
will use Eq. (\ref{epqs2}) with the correspondingly replaced value
of ${\bar{f}}(\mc,\nc)$. Hence stronger entropic bounds of
state-independent form will be obtained. But estimates of the
above kind can initially be posed only for pure states. Suppose
that the relation (\ref{epqs2}) with given ${\bar{f}}(\mc,\nc)$
holds for all the pure state. Then it is still valid for impure
states as well. The justification is based on the Minkowski
inequality (for details, see the proof of proposition 3 in Ref.
\cite{rast102}). So, entropic uncertainty relations with improved
value of ${\bar{f}}(\mc,\nc)$ do also hold for any mixed state. It
is important because many of generalized entropies do not enjoy
the concavity, including the R\'{e}nyi $\alpha$-entropy of order
$\alpha>1$. Before an analysis of the number-phase case, we
consider several examples on base of the result (\ref{trun}).

\section{Examples of uncertainty relations in terms of $(\alpha,s)$-entropies}\label{sc3}

In this section, three interesting examples of uncertainty
relations in terms of the unified entropies are presented. First,
we consider a pair of complementary observables for a $N$-level
system. Second, the angle and the angular momentum are examined.
For the angle, we take discretized probability distribution with
respect to some partition of range of the angle. Finally, we
discuss extremal unravelings of two quantum channels.

{\it Example 1.}
Let complex amplitudes $\tilde{c}_{k}$ and $c_{l}$ be connected by the
discrete Fourier transform
\begin{equation}
\tilde{c}_{k}= \frac{1}{\sqrt{N}}{\>}\sum\nolimits_{l=1}^{N} e^{2\pi{i}k{l}/N}{\>} c_{l}
\ . \label{disf}
\end{equation}
The corresponding probabilities are written as
$p_{k}=|\tilde{c}_{k}|^{2}$ and $q_{l}=|c_{l}|^{2}$. The
transformation (\ref{disf}) is related to a pair of complementary
observables for a $N$-level system \cite{kraus87}. It follows from
$\|{\mathsf{\tilde{c}}}\|_{2}=\|\cp\|_{2}$ and (\ref{suppos}) that
\begin{equation}
\|\cp\|_{a}\leq\left(\frac{1}{\sqrt{N}}\right)^{(2-b)/b}\|{\mathsf{\tilde{c}}}\|_{b}
\ , {\ }\qquad
\|{\mathsf{\tilde{c}}}\|_{a}\leq\left(\frac{1}{\sqrt{N}}\right)^{(2-b)/b}\|\cp\|_{b}
\ , \label{bacb}
\end{equation}
where $1/a+1/b=1$ and $1<b<2$. Squaring these inequalities and
putting $\alpha=a/2$ and $\beta=b/2$, we then obtain
\begin{equation}
\|\qp\|_{\alpha}\leq\left(\frac{1}{N}\right)^{(1-\beta)/\beta}\|\pq\|_{\beta}
\ , {\ }\qquad
{\,} \|\pq\|_{\alpha}\leq\left(\frac{1}{N}\right)^{(1-\beta)/\beta}\|\qp\|_{\beta}
\ . \label{bacb1}
\end{equation}
In other words, we have ${\bar{f}}^{2}=1/N$ here. Using
$(\alpha,s)$-entropies, the uncertainty relation of a
state-independent form are therefore expressed as
\begin{equation}
E_{\alpha}^{(s)}(\pq)+E_{\beta}^{(t)}(\qp)\geq
\frac{1}{\nu}{\>}\ln_{\mu}\bigl(N^{\nu}\bigr)
\ , \label{trune1}
\end{equation}
where $1/\alpha+1/\beta=2$, $st>0$ or $s=t=0$, and the notation
(\ref{nupos}) is assumed. The relation (\ref{trune1}) extends
the previous relations given for both the  R\'{e}nyi
\cite{birula3} and Tsallis entropies \cite{rast104}.

{\it Example 2.}
Let $\{\varphi_{k}\}$ be a partition of the angular interval. The
maximal size $\Delta\varphi:=\max\Delta\varphi_{k}$ of bins
$\Delta\varphi_{k}=\varphi_{k+1}-\varphi_{k}$ is less than $2\pi$.
For given partition, the probability $p_{k}$ to find the angle in
$k$th bin is given by Eq. (\ref{rbin}) with appropriate changes.
Another probability distributions contains $q_{l}=|c_{l}|^{2}$.
Here the coefficients $c_{l}$'s are related to the decomposition
of $\Psi(\varphi)$ with respect to the eigenstates of
$z$-component of the angular momentum, namely
\begin{equation}
\Psi(\varphi)=\frac{1}{\sqrt{2\pi}}\sum_{l=-\infty}^{+\infty}c_{l}{\,}e^{{i}l\varphi}
\ . \label{pcll}
\end{equation}
The Young-Hausdorff inequalities are then written as (see, e.g.,
section 8.17 in Ref. \cite{hardy})
\begin{equation}
\|\Psi\|_a\leq\left(\frac{1}{\sqrt{2\pi}}\right)^{(2-b)/b} \|\cp\|_{b}
\ , {\ }\qquad
\|\cp\|_{a}\leq\left(\frac{1}{\sqrt{2\pi}}\right)^{(2-b)/b} \|\Psi\|_{b}
\ , \label{yona}
\end{equation}
where $1/a+1/b=1$ and $a>2>b$. Applying theorem 192 of Ref.
\cite{hardy} for integral means, we further have
\begin{equation}
\left(\frac{1}{\Delta\varphi_{k}}\int_{\varphi_{k}}^{\varphi_{k+1}}
d\varphi {\,}|\Psi(\varphi)|^{2}\right)^{\alpha}
\left\{
\begin{array}{cc}
\leq, & \alpha>1 \\
\geq, & \alpha<1
\end{array}
\right\}
\frac{1}{\Delta\varphi_{k}}\int_{\varphi_{k}}^{\varphi_{k+1}}
d\varphi{\,}|\Psi(\varphi)|^{2\alpha}
\ . \label{canol}
\end{equation}
Assuming now $\alpha>1>\beta$, we obtain the inequalities
\begin{equation}
\Delta\varphi^{(1-\alpha)/\alpha}\|\pq\|_{\alpha}\leq\|\Psi\|_{a}^{2}
\ , {\ }\qquad
\|\Psi\|_{b}^{2}\leq\Delta\varphi^{(1-\beta)/\beta}\|\pq\|_{\beta}
\ , \label{cor12}
\end{equation}
To derive (\ref{cor12}), we respectively take the inequalities
(\ref{canol}) for $\alpha>1$ and $\beta<1$, sum them with respect
to $k$ and raise the sums to the powers $1/\alpha$ and $1/\beta$.
Combining Eqs. (\ref{cor12}) with the squared Young-Hausdorff
inequalities finally gives
\begin{equation}
\|\pq\|_{\alpha}\leq\left(\frac{\Delta\varphi}{2\pi}\right)^{(1-\beta)/\beta}\|\qp\|_{\beta}
\ , {\ }\qquad
\|\qp\|_{\alpha}\leq\left(\frac{\Delta\varphi}{2\pi}\right)^{(1-\beta)/\beta}\|\pq\|_{\beta}
\ , \label{comb}
\end{equation}
in view of $\|\qp\|_{\beta}=\|\cp\|_{b}^{2}$ and
$(\alpha-1)/\alpha=(1-\beta)/\beta$. In terms of
$(\alpha,s)$-entropies, we then write down
\begin{equation}
E_{\alpha}^{(s)}(\pq)+E_{\beta}^{(t)}(\qp)\geq
\frac{1}{\nu}{\>}\ln_{\mu}\!\left\{\left(\frac{2\pi}{\Delta\varphi}\right)^{\!\nu}\right\}
\ , \label{trune2}
\end{equation}
where $1/\alpha+1/\beta=2$, $st>0$ or $s=t=0$. Particular cases of
Eq. (\ref{trune2}) in terms of the R\'{e}nyi and Tsallis
entropies were derived in Refs. \cite{birula3} and \cite{rast104},
respectively.

For continuous distribution with the probability density function
$W(\varphi)=|\Psi(\varphi)|^{2}$, the unified $(\alpha,s)$-entropy
is expressed by Eqs. (\ref{cddfs}) and (\ref{cddf0}) with
$W(\varphi)$ instead of $P(\theta)$. Note again that only
discretized distributions are actually related to real
experiments. So, entropic inequalities with continuous
distributions are rather interesting as some limiting varieties of
inequalities with discrete distributions, when the size of bins
tends to zero. First, we square both the inequalities of Eq.
(\ref{yona}) and obtain
\begin{equation}
\|W\|_{\alpha}\leq\left(\frac{1}{2\pi}\right)^{(1-\beta)/\beta} \|\qp\|_{\beta}
\ , {\ }\qquad
\|\qp\|_{\alpha}\leq\left(\frac{1}{2\pi}\right)^{(1-\beta)/\beta} \|W\|_{\beta}
\ , \label{yona1}
\end{equation}
where $1/\alpha+1/\beta=2$ and $\alpha>1>\beta$. In the R\'{e}nyi
case, when $s=0$, the uncertainty relation is posed as
\begin{equation}
R_{\alpha}(W)+R_{\beta}(\qp)\geq\ln(2\pi)
\ , \label{trune0}
\end{equation}
for all $\alpha$ and $\beta$ that obey $1/\alpha+1/\beta=2$. The
derivation is simple. Taking the logarithm of the first inequality
of Eq. (\ref{yona1}), one gets
\begin{equation}
\frac{1-\alpha}{\alpha}{\ }\frac{\beta}{1-\beta}{\ }R_{\alpha}(W)\leq
-\ln(2\pi)+R_{\beta}(\qp)
\ . \label{yabb}
\end{equation}
To obtain Eq. (\ref{trune0}) for $\alpha>\beta$, we merely notice
that the multiplier of $R_{\alpha}(W)$ is equal to $(-1)$. In a
similar manner, we resolve the case, when R\'{e}nyi's entropy of
the distribution $\qp$ has larger order.

For $s\neq0$, we cannot generally apply the statement of Lemma
\ref{lemun} in view of the following reasons. Within
minimization, the restriction to the domain (\ref{dodf})
is crucial. If one of the variables $\xi$ and $\zeta$ takes values
from both the ranges $(0;1)$ and $(1;+\infty)$ then the reasons
from the proof of Lemma \ref{lemun} fail. So we should restrict
our consideration to those functions $W(\varphi)$ that obey
$\|W\|_{\alpha}\leq\|W\|_{1}=1$ for $\alpha>1$ and
$\|W\|_{\beta}\geq\|W\|_{1}=1$ for $\beta<1$. For instance, the
last conditions are clearly satisfied, when $W(\varphi)\leq1$ for
all $\varphi\in[0;2\pi]$. Combining Eq. (\ref{yona1}) with
Lemma \ref{mthm}, we then obtain
\begin{equation}
\eas(W)+\ebt(\qp)\geq \frac{1}{\nu}
{\>}\ln_{\mu}\bigl\{(2\pi)^{\nu}\bigr\}
\ , \label{npuen}
\end{equation}
where $1/\alpha+1/\beta=2$, $st>0$ or $s=t=0$. Values of
$(\alpha,s)$-entropies become negative, when the above conditions
on norm-like functionals do not hold for given probability density
function $W(\varphi)$. In such a case, entropies of the continuous
distribution hardly have a physical meaning. Instead of such
entropies, we then consider entropies of a discretized angular
distribution and obtain Eq. (\ref{trune2}).

{\it Example 3.}
Changes of states in quantum theory are generally represented by
linear maps \cite{nielsen}. Maps of such a kind must be
completely positive and, when describe deterministic processes,
trace-preserving as well. Each completely positive map $\Phi$ can
be written in the operator-sum representation. For each operator
$\ax$ on the input Hilbert space, we have
\begin{equation}
\Phi(\ax)=\sum\nolimits_{j}\am_{j}{\,}\ax{\,}\am_{j}^{\dagger}
\ , \label{oper3}
\end{equation}
where Kraus operators $\am_{j}$ map the input Hilbert space to the
output one \cite{watrous1,nielsen}. The preservation of the trace
implies that $\tr\bigl(\Phi(\ax)\bigr)=\tr(\ax)$ for all $\ax$,
whence
\begin{equation}
\sum\nolimits_{j}\am_{j}^{\dagger}{\,}\am_{j}=\pen
\ . \label{supn}
\end{equation}
Trace-preserving completely positive maps are usually called
''quantum channels'' \cite{nielsen}. For given quantum channel,
much many operator-sum representations exist. It the sets
$\ac=\{\am_{j}\}$ and $\bc=\{\bn_{i}\}$ represent the same quantum
channel then
\begin{equation}
\bn_{i}=\sum\nolimits_{j} \am_j{\,}u_{ji}
\ , \label{eqvun}
\end{equation}
where the matrix $\um=[[u_{ij}]]$ is unitary
\cite{watrous1,nielsen}. Each set $\ac=\{\am_{i}\}$ that does obey
(\ref{oper3}) will be named an ''unraveling'' of the channel
$\Phi$. This terminology is due to Carmichael \cite{carm} who used
this word for a representation of the master equation.
Applications of the master equation and its unravelings in quantum
control are reviewed in Ref. \cite{wm10}. It is of interest that
there exist so-called ''extremal unravelings''
\cite{breslin,ilichev03}. In Ref. \cite{rast11a}, we examine those
unravelings of a channel that are extremal with respect to
$(\alpha,s)$-entropies. For given density matrix $\rho$ and
unraveling $\ac=\{\am_{j}\}$, we define the matrix
\begin{equation}
\pim(\ac|\roh):=\bigl[\bigl[\langle\am_{i}\sqrt{\roh}{\,},\am_{j}\sqrt{\roh}\rangle_{\rm{hs}}\bigr]\bigr]
=\bigl[\bigl[\tr(\am_{i}^{\dagger}\am_{j}\roh)\bigr]\bigr]
\ . \label{pimdef}
\end{equation}
The matrix $\pim(\ac|\roh)$ is Hermitian, its diagonal element
$p_{i}=\tr(\am_{i}^{\dagger}\am_{i}\roh)$ gives the $i$th effect
probability. Then the entropy $\eas(\ac|\rho)$ is defined by Eq.
(\ref{undef}) with these probabilities. Note that matrices of the
form (\ref{pimdef}) were used by Lindblad \cite{glind79} to
introduce the entropic quantity, which is known in quantum
information as the entropy exchange. If $\roh$ is taken to be
completely mixed, then the entropy exchange coincides with the map
entropy of $\Phi$ \cite{rfz10}. The latter is another
characteristic of the channel related to its Jamio{\l}kowski--Choi
representation. Due to Eq. (\ref{eqvun}), the matrices
$\pim(\ac|\rho)$ and $\pim(\bc|\rho)$ are unitarily similar
\cite{rast104}, namely
\begin{equation}
\pim(\bc|\roh)=\um^{\dagger}{\,}\pim(\ac|\roh){\,}\um
\ . \label{usab}
\end{equation}
So, for given unraveling $\ac=\{\am_i\}$ we build $\pim(\ac|\rho)$
and diagonalize it as
$\vm^{\dagger}{\,}\pim(\ac|\roh){\,}\vm={\mathrm{diag}}(\lambda_{1},\lambda_{2},\ldots)$.
Using this unitary matrix $\vm=[[v_{ij}]]$, we then define a
specific unraveling $\ac_{\rho}^{(ex)}$ such that
\begin{equation}
\am_{i}^{(ex)}:=\sum\nolimits_{j} \am_j{\,} v_{ji}
\ . \label{excalc}
\end{equation}
In Ref. \cite{rast11a}, we have shown the extremality of
$\ac_{\rho}^{(ex)}=\bigl\{\am_{i}^{(ex)}\bigr\}$ with respect to
almost all of the $(\alpha,s)$-entropies. That is, each unraveling
$\ac$ of the channel $\Phi$ obeys
\begin{equation}
\eas(\ac|\roh)\geq\eas(\ac_{\rho}^{(ex)}|\roh)
\ , \label{uneex}
\end{equation}
where $\alpha>0$ and $s\neq0$. In the case of R\'{e}nyi's
$\alpha$-entropy, when $s=0$, the unraveling with Kraus operators
(\ref{excalc}) enjoys the extremality property (\ref{uneex}) only
for $0<\alpha<1$ \cite{rast104}. In the case $\alpha=1$ we deal
with the Shannon entropy, for which the property (\ref{uneex}) was
considered in Refs. \cite{breslin,ilichev03}. We do not consider
here unravelings that are extremal with respect to R\'{e}nyi's
entropies of order $\alpha>1$. Such unravelings differ from the
unraveling with elements (\ref{excalc}) and may also depend on
$\alpha$ in general. In view of the lower bound (\ref{uneex}), we
properly pose entropic uncertainty relations for extremal
unravelings of two quantum channels. Suppose that the input
density matrix $\roh$ is given. Let $\ac_{\rho}^{(ex)}$ and
$\bc_{\rho}^{(ex)}$ be extremal unravelings of the quantum
channels $\Phi_{\ac}$ and $\Phi_{\bc}$, respectively. It follows
from Eq. (\ref{trun}) that
\begin{equation}
\eas(\ac_{\rho}^{(ex)}|\roh)+\ebt(\bc_{\rho}^{(ex)}|\roh)
\geq\frac{1}{\nu}\ln_{\mu}\Bigl\{g\bigl(\ac_{\rho}^{(ex)},\bc_{\rho}^{(ex)}|\roh\bigr)^{-2\nu}\Bigr\}
\ , \label{ennun}
\end{equation}
where $1/\alpha+1/\beta=2$, $st>0$, the $\mu$ and $\nu$ are put by
Eq. (\ref{nupos}). For the particular case of Tsallis' entropies,
this relation was derived in Ref. \cite{rast104}. The uncertainty
relation (\ref{ennun}) remains valid for the Shannon entropies,
when $\alpha=\beta=1$. Except for the latter, either of orders
$\alpha$ and $\beta$ of the two R\'{e}nyi entropies is larger than
one. Hence, the analog of Eq. (\ref{ennun}) with R\'{e}nyi's
entropies will include an unknown extremal unraveling, which
cannot be found from Eq. (\ref{excalc}). So we refrain from
presenting such relations here.

\section{Entropic formulation of number-phase uncertainty relations}\label{sc4}

In this section, number-phase uncertainty relations in terms of
$(\alpha,s)$-entropies are derived within the Pegg--Barnett
formalism. Since the limit is involved here, we should consider
Eq. (\ref{epqs1}) again and take $N\to\infty$ in its
state-independent variety. As a result, a similar inequality with
the continuous distribution with respect to $\theta$ is obtained.
Using this inequality, we derive several number-phase uncertainty
relations of a state-independent form in terms of
$(\alpha,s)$-entropies. Most of them are written for discrete
phase distributions. The case, which allows relations with
continuous phase distribution, is considered as well.

\subsection{Inequalities and entropic relations of ''number-phase'' type for finite $N$}\label{s41}

Let $|\psi_{N+1}\rangle\in\hh_{N+1}$ be a normalized vector. We
can decompose it with respect to the bases $\{|n\rangle\}$ and
$\{|\theta_m\rangle\}$. Using the corresponding resolutions of the
identity, one gives
\begin{equation}
\pen|\psi_{N+1}\rangle=\sum\nolimits_{n=0}^{N}x_{n}|n\rangle=
\sum\nolimits_{m=0}^{N}y_{m}|\theta_{m}\rangle
\ . \label{twrs}
\end{equation}
The complex numbers $x_n=\langle{n}|\psi_{N+1}\rangle$ and
$y_m=\langle\theta_{m}|\psi_{N+1}\rangle$ form the two
$(N+1)$-tuples $\xp$ and $\yp$, respectively. These tuples satisfy
$\|\xp\|_{2}^{2}=\|\yp\|_{2}^{2}=\langle\psi_{N+1}|\psi_{N+1}\rangle=1$.
So, we can apply the Riesz theorem. Due to the orthonormality, we at
once obtain
\begin{equation}
x_{n}=\sum_{m=0}^{N}\langle{n}|\theta_m\rangle{\,}y_{m}=
\sum_{m=0}^{N}\frac{e^{i{n}\theta_{m}}}{\sqrt{N+1}}{\ }y_{m}
\ , \qquad
y_{m}=\sum_{n=0}^{N}\langle\theta_{m}|n\rangle{\,}x_{n}=
\sum_{n=0}^{N}\frac{e^{-i{n}\theta_{m}}}{\sqrt{N+1}}{\ }x_{n}
\ . \label{tranxy}
\end{equation}
In both the cases, all elements of the transformation matrix $\ttm$ by
size $(N+1)\times(N+1)$ have the same absolute value equal to
$(N+1)^{-1/2}$. Substituting $\eta=(N+1)^{-1/2}$ into Eq.
(\ref{suppos}), for $1/a+1/b=1$ and $1<b<2$ there holds
\begin{equation}
\|\xp\|_{a}\leq\left(\frac{1}{\sqrt{N+1}}\right)^{(2-b)/b}\|\yp\|_{b}
\ , \qquad
\|\yp\|_{a}\leq\left(\frac{1}{\sqrt{N+1}}\right)^{(2-b)/b}\|\xp\|_{b}
\ . \label{xpyp}
\end{equation}
In terms of probabilities
$p_{m}=|y_{m}|^{2}=|\langle\theta_{m}|\psi_{N+1}\rangle|^{2}$ and
$q_{n}=|x_{n}|^{2}=|\langle{n}|\psi_{N+1}\rangle|^{2}$, the squared
relations (\ref{xpyp}) are rewritten as
\begin{equation}
\|\qp\|_{\alpha}\leq\left(\frac{1}{N+1}\right)^{(1-\beta)/\beta}\|\pq\|_{\beta}
\ , \qquad
\|\pq\|_{\alpha}\leq\left(\frac{1}{N+1}\right)^{(1-\beta)/\beta}\|\qp\|_{\beta}
\ , \label{xpyp1}
\end{equation}
where $1/\alpha+1/\beta=2$ and $\alpha>1>\beta$. Due to Lemma
\ref{lemun}, the inequalities (\ref{xpyp1}) lead to the bound
(\ref{trune1}) with $(N+1)$ instead of $N$. So, this is a
unified-entropy uncertainty relation for two mutually unbiased
bases. To obtain an uncertainty relation for the phase and number
operators, we must consider the limit $N\to\infty$. However, the
right-hand side of (\ref{trune1}) is divergent in this limit. One
of ways to formulate a meaningful statement was proposed in Ref.
\cite{gvb95}. This approach allows to obtain uncertainty relations
in terms of the Shannon entropies \cite{gvb95,hatami}. For
generalized entropies, however, other methods should be used.

\subsection{Relations between norm-like functionals in the limit $N\to\infty$}\label{s42}

The key idea of our approach is to take the desired limit directly
in relations between norms. So, we shall now rewrite the
inequalities (\ref{xpyp}) in an appropriate way. Below we will
omit any subscript for those objects that stand for taking
$N\to\infty$. To given $(N+1)$-tuple $\xp=\{x_{n}\}$, we assign the
function $F_{N+1}(\theta)$ of variable $\theta$, namely
\begin{equation}
F_{N+1}(\theta):=\frac{1}{\sqrt{2\pi}}{\,}\sum_{n=0}^{N} e^{-in\theta}{\,}x_{n}
\ . \label{fsdf}
\end{equation}
When $N\to\infty$, the infinite series of Eq. (\ref{fsdf}) defines
some $2\pi$-periodic function $F(\theta)$. In view of the relation
\begin{equation}
\frac{1}{2\pi}\int\nolimits_{0}^{2\pi}\exp[i(m-n)\theta]{\,}d\theta=\delta_{mn}
\   , \label{rorel}
\end{equation}
there holds
\begin{equation}
\|F_{N+1}\|_{2}^{2}=\int\nolimits_{0}^{2\pi}|F_{N+1}(\theta)|^{2}{\,}d\theta
=\sum\nolimits_{n=0}^{N}|x_{n}|^{2}=1
\ . \label{fsnm2}
\end{equation}
By virtue of Eq. (\ref{tranxy}), the elements of the tuple $\yp$
are then represented as
\begin{equation}
y_{m}=\langle\theta_{m}|\psi_{N+1}\rangle=
\sqrt{\frac{2\pi}{N+1}}{\ }F_{N+1}(\theta_{m})
\ . \label{fsym}
\end{equation}
The right-hand side of the first inequality in Eq. (\ref{xpyp}) is
therefore rewritten as
\begin{align}
 & \left(\frac{1}{\sqrt{N+1}}\right)^{(2-b)/b}\sqrt{\frac{2\pi}{N+1}}{\>}
\left(\frac{N+1}{2\pi}\right)^{1/b}\left({\,}\sum_{m=0}^{N} |F_{N+1}(\theta_{m})|^b\Delta\theta\right)^{1/b}=
\nonumber\\
 & \left(\frac{1}{\sqrt{2\pi}}\right)^{(2-b)/b}
\left({\,}\sum_{m=0}^{N} |F_{N+1}(\theta_{m})|^b\Delta\theta\right)^{1/b},
\label{fsxx}
\end{align}
where $\Delta\theta=2\pi/(N+1)$. The set
$\{\theta_{0},\theta_{1},\ldots,\theta_{N}\}$ with
$\theta_{N}=\theta_0+2\pi{N}/(N+1)$ is a partition of the
half-open interval $[\theta_{0};\theta_{0}+2\pi)$. For brevity, we
take $\theta_{0}=0$. In the limit $N\to\infty$, the sum in the
right-hand side of Eq. (\ref{fsxx}) is clearly converted into a
Riemann--Darboux integral of periodic function $F(\theta)$. Hence
the first inequality in Eq. (\ref{xpyp}) becomes
\begin{equation}
\|\xp\|_{a}\leq\left(\frac{1}{\sqrt{2\pi}}\right)^{(2-b)/b}
\left(\int\nolimits_{0}^{2\pi}|F(\theta)|^b{\,}d\theta\right)^{1/b}
=\left(\frac{1}{\sqrt{2\pi}}\right)^{(2-b)/b}\|F\|_{b}
\ . \label{limpxb}
\end{equation}
In a similar manner, the second inequality in Eq. (\ref{xpyp}) is
merely converted into
\begin{equation}
\left(\int\nolimits_{0}^{2\pi}|F(\theta)|^a{\,}d\theta\right)^{1/a}=
\|F\|_{a}\leq\left(\frac{1}{\sqrt{2\pi}}\right)^{(2-b)/b}\|\xp\|_{b}
\ . \label{limpxa}
\end{equation}
Since $\|F_{N+1}\|_{2}=1$ by construction, one enjoys $\|F\|_{2}=1$
as well. For pure state
$|\psi\rangle=\sum_{n=0}^{\infty}{x_{n}|n\rangle}$ (the state
(\ref{twrs}) in the limit $N\to\infty$), the formulas (\ref{ppdf})
and (\ref{fsym}) give
\begin{equation}
P(\theta)=\underset{N\to\infty}{\lim}
\frac{N+1}{2\pi}{\>}|\langle\theta_m|\psi_{N+1}\rangle|^{2}=|F(\theta)|^{2}
\ . \label{ppdf1}
\end{equation}
The equality $\|F\|_{2}=1$ leads to
$\int_{0}^{2\pi}P(\theta){\,}d\theta=1$. In terms of parameters
$\alpha=a/2$ and $\beta=b/2$, we also have
$\|F\|_{a}^{2}=\|P\|_{\alpha}$ and $\|F\|_{b}^{2}=\|P\|_{\beta}$.
Squaring Eqs. (\ref{limpxb}) and (\ref{limpxa}), we transform them
into the desired relations between norm-like functionals, namely
\begin{equation}
\|\qp\|_{\alpha}\leq\left(\frac{1}{2\pi}\right)^{(1-\beta)/\beta}\|P\|_{\beta}
\ , \qquad
\|P\|_{\alpha}\leq\left(\frac{1}{2\pi}\right)^{(1-\beta)/\beta}\|\qp\|_{\beta}
\ . \label{xpyp8}
\end{equation}
These inequalities are analogous to the inequalities
(\ref{xpyp1}). We shall now use them for deriving number-phase
uncertainty relations in terms of unified entropies.

\subsection{Number-phase uncertainty relations for arbitrary states}\label{s43}

Using the method of Refs. \cite{rast104,rast105}, one can obtain
an uncertainty relation with some discretization of probability
density function $P(\theta)$. For a partition $\{\vata_{m}\}$ of the
interval $[0;2\pi]$, we introduce probabilities by Eq.
(\ref{rbin}). Due to theorem 192 of the book \cite{hardy} for
integral means, we have
\begin{equation}
\frac{r_{m}^{\alpha}}{\Delta\vata_{m}^{\alpha}}=\left(\frac{1}{\Delta\vata_{m}}{\>}
\int\nolimits_{\vata_{m}}^{\vata_{m+1}} P(\theta){\,}d\theta\right)^{\!\alpha}
{\>}\left\{
\begin{array}{cc}
\leq, & \alpha>1 \\
\geq, & \alpha<1
\end{array}
\right\}
{\>}\frac{1}{\Delta\vata_{m}}{\>}
\int\nolimits_{\vata_m}^{\vata_{m+1}} P(\theta)^{\alpha}{\,}d\theta
\ , \label{conva}
\end{equation}
where $\Delta\vata_{m}=\vata_{m+1}-\vata_{m}$. Summing Eq.
(\ref{conva}) with respect to $m$ and putting
$\Delta\vata=\max\Delta\vata_{m}$, we get
\begin{equation}
\Delta\vata^{1-\alpha}{\,}\|\rp\|_{\alpha}^{\alpha}=\Delta\vata^{1-\alpha}{\,}\sum\nolimits_{m} r_{m}^{\alpha}
{\>}\left\{
\begin{array}{cc}
\leq, & \alpha>1 \\
\geq, & \alpha<1
\end{array}
\right\}
{\>}\int\nolimits_{0}^{2\pi} P(\theta)^{\alpha}{\,}d\theta=\|P\|_{\alpha}^{\alpha}
\ . \label{convaa}
\end{equation}
Using the last relation, we convert the inequalities (\ref{xpyp8})
to the form
\begin{equation}
\|\qp\|_{\alpha}\leq\left(\frac{\Delta\vata}{2\pi}\right)^{(1-\beta)/\beta}\|\rp\|_{\beta}
\ , \qquad
\|\rp\|_{\alpha}\leq\left(\frac{\Delta\vata}{2\pi}\right)^{(1-\beta)/\beta}\|\qp\|_{\beta}
\ , \label{xpypr}
\end{equation}
where $1/\alpha+1/\beta=2$ and $\alpha>1>\beta$. Combining the
statement of Lemma \ref{lemun} with Eq. (\ref{xpypr}), the
uncertainty relations are written in the following way.

\newtheorem{t51}[t32]{Theorem}
\begin{t51}\label{vatat}
Let $\{\vata_{m}\}$ be a partition of the interval $[0;2\pi]$. The
unified entropies of the corresponding probability distribution
$\rp=\{r_{m}\}$, defined by Eq. (\ref{rbin}), and the number
probability distribution $\qp=\{q_{n}\}$ satisfy
\begin{equation}
\eas(\rp)+\ebt(\qp)\geq \frac{1}{\nu}
{\>}\ln_{\mu}\left\{\left(\frac{2\pi}{\Delta\vata}\right)^{\!\nu{\>}}\right\}
\ . \label{pnued}
\end{equation}
Here $st>0$ or $s=t=0$, positive $\alpha$ and $\beta$ obey
$1/\alpha+1/\beta=2$, the parameters $\mu$ and $\nu$ are defined
by Eq. (\ref{nupos}).
\end{t51}

Since all the bins are strictly less than $2\pi$, this bound on
the sum of two entropies is nontrivial. For the particular cases
$s=t=0$ and $s=t=1$, the above uncertainty relation is reduced to
the entropic bounds
\begin{equation}
R_{\alpha}(\rp)+R_{\beta}(\qp)\geq
\ln\bigl(2\pi/\Delta\vata\bigr)
\ , \qquad
H_{\alpha}(\rp)+H_{\beta}(\qp)\geq
\ln_{\mu}\bigl(2\pi/\Delta\vata\bigr)
\ . \label{rh01}
\end{equation}
These bounds are number-phase relations in terms of the R\'{e}nyi
and Tsallis entropies, respectively. Entropic inequalities,
involving the continuous distribution $P(\theta)$, are also of
interest. For the R\'{e}nyi case, number-phase uncertainty
relations are posed similarly to Eq. (\ref{trune0}).

\newtheorem{t52}[t32]{Theorem}
\begin{t52}\label{reeth}
Under the condition $1/\alpha+1/\beta=2$, the R\'{e}nyi entropies
of the phase probability density $P(\theta)$ and the number
probability distribution $\qp=\{q_{n}\}$ satisfy
\begin{equation}
R_{\alpha}(P)+R_{\beta}(\qp)\geq\ln(2\pi)
\ . \label{rnuen}
\end{equation}
\end{t52}

For $\alpha=\beta=1$, the inequality (\ref{rnuen}) gives the lower
bound $\ln(2\pi)$ on the sum of the corresponding Shannon
entropies. This bound was presented in Refs. \cite{gvb95,jos01}.
Some doubts with the entropies of continuous distributions were
mentioned above. Nevertheless, if we restrict our consideration to
probability density functions, obeying $P(\theta)\leq1$ for all
$\theta\in[0;2\pi]$, then the entropic functionals (\ref{cddfs})
and (\ref{cddf0}) are positive-valued. Hence the entropic uncertainty
relations are written as
\begin{equation}
\eas(P)+\ebt\bigl(\qp\bigr)\geq \frac{1}{\nu}
{\>}\ln_{\mu}\bigl\{(2\pi)^{\nu}\bigr\}
\ . \label{pnuen}
\end{equation}
As above, $st>0$, the parameters $\alpha>0$ and $\beta>0$ satisfy
$1/\alpha+1/\beta=2$, the parameters $\mu$ and $\nu$ are defined
by Eq. (\ref{nupos}).

\subsection{Number-phase uncertainty relations for multiphoton coherent states}\label{s44}

Consider coherent states $|z\rangle$ with the mean photon number
$|z|^{2}\gg1$. As usual, we approximate the corresponding phase and
number distributions by the Gaussian functions (\ref{gampa}) and
(\ref{nampa}). For the Fourier transform, the exact values of
$(b,a)$-norms were found by Beckner \cite{beck}. In the notation
(\ref{tipw}), the relations between norms are such that
\begin{equation}
\|\tiw\|_{a}\leq{c}_{b}{\,}\|\tip\|_{b}
\ , \qquad
\|\tip\|_{a}\leq{c}_{b}{\,}\|\tiw\|_{b}
\ , \qquad
c_{b}^{2}=\left(\frac{b}{2\pi}\right)^{1/b}\left(\frac{2\pi}{a}\right)^{1/a}
\ , \label{tpwc}
\end{equation}
where $1/a+1/b=1$ and $a>2>b$. Using
$\|\tiw\|_{a}^{2}=\|\wiw\|_{\alpha}$ and
$\|\tip\|_{b}^{2}=\|\wip\|_{\beta}$ and squaring Eq. (\ref{tpwc}),
we obtain
\begin{equation}
\|\wiw\|_{\alpha}\leq{C}_{\beta}{\,}\|\wip\|_{\beta}
\ , \qquad
\|\wip\|_{\alpha}\leq{C}_{\beta}{\,}\|\wiw\|_{\beta}
\ , \qquad
C_{\beta}=\left(\frac{\beta}{\pi}\right)^{1/(2\beta)}\biggl(\frac{\pi}{\alpha}\biggr)^{1/(2\alpha)}
\ . \label{tpws}
\end{equation}
As above, we put $\alpha=a/2$ and $\beta=b/2$ so that
$1/\alpha+1/\beta=2$ and $\alpha>1>\beta$. Let us introduce the
probabilities $\tir_{m}$ by Eq. (\ref{rbin}) with $\wip(\theta)$
and
\begin{equation}
\tiq_{n}=\int\nolimits_{n}^{n+1} \wiw(y){\,}dy
\ , \label{qbin}
\end{equation}
for integer $n$. Similarly to Eq. (\ref{xpypr}), we derive from
Eq. (\ref{tpws}) that
\begin{equation}
\|\wqp\|_{\alpha}\leq\left(\frac{\Delta\vata}{\pi}\right)^{(1-\beta)/\beta}
\frac{\beta^{1/(2\beta)}}{\alpha^{1/(2\alpha)}}{\ }\|\wrp\|_{\beta}
\ , \qquad
\|\wrp\|_{\alpha}\leq\left(\frac{\Delta\vata}{\pi}\right)^{(1-\beta)/\beta}
\frac{\beta^{1/(2\beta)}}{\alpha^{1/(2\alpha)}}{\ }\|\wqp\|_{\beta}
\ . \label{tpww}
\end{equation}
Combining these inequalities with the statement of Lemma
\ref{lemun} finally yields the entropic uncertainty relation
\begin{equation}
\eas(\wrp)+\ebt(\wqp)\geq \frac{1}{\nu}
{\>}\ln_{\mu}\left\{\Bigl(\frac{\varkappa\pi}{\Delta\vata}\Bigr)^{\nu}\right\}
\ . \label{npmph}
\end{equation}
Here $st>0$ or $s=t=0$, the parameters $\alpha$ and $\beta$ obey
$1/\alpha+1/\beta=2$, and square of the factor $\varkappa$ is put
by
\begin{equation}
\varkappa^{2}=\alpha^{1/(\alpha-1)}\beta^{1/(\beta-1)}
=\exp_{\alpha}(1/\alpha){\,}\exp_{\beta}(1/\beta)
\ . \label{mpmpf}
\end{equation}
Under the condition $1/\alpha+1/\beta=2$, the right-hand side
of Eq. (\ref{mpmpf}) monotonically increases with
$\beta\in(1/2;1)$. The factor increases from $\varkappa=2$ for
$\beta=1/2$ right up to $\varkappa=e$ for $\beta=1$. For the
former, the multiphoton-state bound (\ref{npmph}) concurs with the
general bound (\ref{pnued}). The distinction between the two
bounds is maximal, when both the $\alpha$ and $\beta$ tend to one.
In this case, the relation in terms of the Shannon entropies reads
\begin{equation}
S(\wrp)+S(\wqp)\geq\ln\left(\frac{e\pi}{\Delta\vata}\right)
\ . \label{npmsh}
\end{equation}
Similar to Eqs. (\ref{trune0}) and (\ref{rnuen}), one also gets an
uncertainty relation in terms of R\'{e}nyi's entropies in the form
\begin{equation}
R_{\alpha}(\wip)+R_{\beta}(\wiw)\geq\ln(\varkappa\pi)
\ , \label{nuenr}
\end{equation}
where $1/\alpha+1/\beta=2$ and integrals of the probability
density functions are involved. For the position and momentum, the
same entropic bound was derived in Ref. \cite{birula3}. For the
Shannon entropies, the relation (\ref{nuenr}) gives the lower
bound $\ln(e\pi)$ previously presented in Refs.
\cite{abe92,gvb95,jos01}.  Note that the relation (\ref{npmph})
with finite resolutions can easily be reformulated for the case of
canonically conjugate position and momentum. We refrain from
presenting the details here.

\section{Conclusions}\label{sc5}

We have considered a formulation of number-phase uncertainty
relations in terms of unified entropies, which include both the
R\'{e}nyi and Tsallis ones as particular cases. For two
generalized measurements, unified-entropy uncertainty relations of
state-dependent as well as state-independent form are derived.
Using Riesz's theorem, nontrivial inequalities between norm-like
functionals of two generated probability distributions were
obtained. The proposed method combines this constraint with a task
of minimizing a certain function of two variables in the
corresponding domain of their acceptable values. Hence we have
stated unified-entropy uncertainty relations which are an
extension of previous bounds in terms of the R\'{e}nyi and Tsallis
entropies. We also gave examples of two complementary observables
in a $N$-level system, angle and angular momentum, and extremal
unravelings of two quantum channels.

To obtain number-phase uncertainty relations, the
infinite-dimensional limit should be treated properly. Using the
Pegg--Barnett formalism and the Riesz theorem, we derived an
inequality for functionals of the corresponding probability
distributions in finite dimensions. Then we have taken the limit
$N\to\infty$ right in this inequality. Hence unified-entropy
uncertainty relations for the number-phase pair are naturally
established. As it was shown, entropic bounds can be improved for
the case of coherent states with large mean number of photons. The
derived uncertainty relations are mainly formulated for phase
distribution discretized with respect to finite experimental
resolutions. Some entropic bounds with the continuous phase
distribution were given as well. A lot of the presented results is
a proper extension of entropic uncertainty relations previously
reported in the literature.

\end{document}